\documentclass[fleqn,10pt]{wlscirep}
\usepackage{graphicx}
\usepackage{braket}
\usepackage{amsmath}
\usepackage{blkarray}
\usepackage{multirow}
\usepackage{subcaption}
\usepackage{nameref}
\usepackage{hyperref}
\usepackage{color}
\usepackage{tikz}

\usepackage[mathscr]{euscript}
\usepackage{calrsfs}
\DeclareMathAlphabet{\pazocal}{OMS}{zplm}{m}{n}

\newcommand{\Hb}{\pazocal{H}}
\newcommand{\Ub}{\pazocal{U}}

\title{Dual Quantum Zeno Superdense Coding}

\author{Fakhar Zaman}
\author{Youngmin Jeong}
\author{Hyundong Shin}
\affil{Department of Electronic Engineering, Kyung Hee University, Yongin-si, 17104 Korea \\ Correspondence and request for materials should be addressed to Y.J and H.S (email: yjeong@khu.ac.kr; hshin@khu.ac.kr)}

\begin{abstract}
Quantum superdense coding enables a sender to encode a two-bit classical message in one qubit using the preshared entanglement.  In this paper, we develop a superdense coding protocol using a \textit{dual quantum Zeno $\left(\text{DQZ}\right)$} gate to take the full advantage of quantum superdense coding from the complete Bell-state analysis. We verify that the DQZ gate allows remote parties to achieve the distinguishability of orthonormal Bell states in a semi-counterfactual manner and the DQZ superdense coding achieves high throughput efficiency as a function of cycle numbers for the Bell-state analyzer.
\end{abstract}

\begin{document}

\flushbottom
\maketitle
%
%
\thispagestyle{empty}


\section*{Introduction}

Quantum superdense coding is a communication protocol where two-bit classical information can be transferred between two spatially separated parties using initially shared entanglement \cite{BW:92:PRL}.  In general, one can encode $2n$ bits of classical information in $n$ qubits from shared $2n$-qubit maximally entangled states. Thus, the pre-shared maximally entangled state can double the classical capacity of the channel. However, the efficiency of the quantum superdense coding depends on the distinguishability of the $2^{2n}$ maximally entangled states.

For a bipartite system, an important challenge in realizing quantum superdense coding is the distinguishability of the Bell states \cite{BW:92:PRL,MWKA:96:PRL}. However, one cannot discriminate between the four Bell states using only linear elements \cite{VY:99:PRA,CN:01:APB,GKRSS:01:PRL}. The use of hyperentanglement \cite{K:97:JMO,BLPK:05:PRL}, the entanglement in multiple degrees of freedom, or an ancillary photon \cite{G:11:PRA,HM:08:PRA} with linear optics made it possible to discriminate between the four orthonormal Bell states with certainty. This ancillary degree of freedom carries no information; rather, it expands the dimensions of the system to enhance the distinguishability of the Bell states. The hyperentangled Bell-state analysis has been demonstrated using the orbital angular momentum \cite{BWK:08:NP} or time \cite{SHKW:06:PRL,WRT:17:PRL} as the ancillary degree of freedom. 
A different approach for the complete Bell-basis measurement has been presented under the asymptotic limits \cite{A:04:PRA}. The authors proposed a two-qubit interaction-free measurement $\left(\text{IFM}\right)$ gate that either the photon collapses back to the initial state or changes its trajectory depends on the absence or presence of the absorptive object, and a controlled NOT (CNOT) gate by chaining multiple two-qubit IFM gates. This make it complex to implement as compared to the simplicity of the task. Later, several schemes for the CNOT gate operation via the IFM gate have been presented \cite{FJP:04:PRA,HM:08:PRA}. However,  these schemes for the IFM CNOT gate sacrifice the throughput of quantum superdense coding. 

In this paper, we design superdense coding with the advantage of complete Bell-state analysis using a \textit{dual quantum Zeno $\left(\text{DQZ}\right)$} gate. The quantum Zeno $\left(\text{QZ}\right)$  effect \textit{alters the decay rate of an unstable system by frequent measurements. It freezes the evolution of the system if the frequency of the repeated measurements is high enough.} \cite{IHBW:90:PRA,PTP:90:PRA}. We demonstrate that the DQZ Bell-state analyzer enables i) remote parties to achieve the semi-counterfactual distinguishability of the Bell states and ii) quantum superdense coding to achieve high throughput efficiency.

In the later half of this section, we will briefly review the IFM and quantum superdense coding using IFM  and QZ CNOT gates \cite{A:04:PRA,HM:08:PRA}. In~\hyperref[sec: Results]{Results}, we initially propose our DQZ gate pursued by the Bell-state analysis. At the end, we  provide the throughput efficiency for DQZ superdense coding as an element of the quantity of cycles for the Bell-state analyzer. In~\hyperref[sec: Discussion]{Discussion}, we briefly conclude our scheme and discuss the semi-counterfactual behavior of our DQZ Bell-state analyzer. In~\hyperref[sec: Methods]{Methods}, we drive the throughput efficiency $R$ [bits/qubit] for superdense coding with IFM and QZ Bell-state analyzers. 

\begin{figure}[t!]
 \centering
\includegraphics[width=0.7\textwidth]{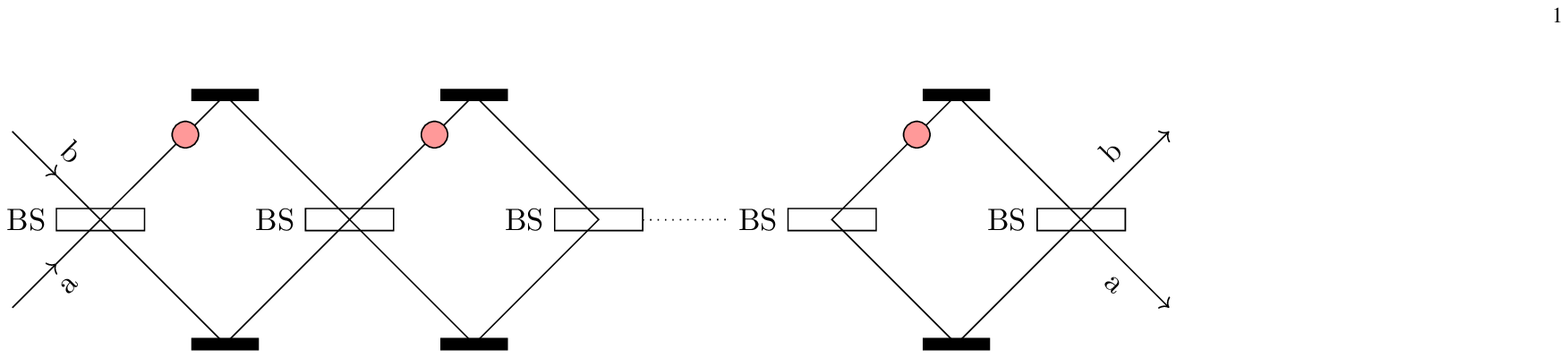}
    \caption{
  \textbf{Interaction-free measurement.} Here BS stands for an unbalanced beam splitter. The output state of the photon after $N$ cycles determines the state of the absorptive object. If there is an absorptive object in path b, the photon ends up in the state $\ket{01}_{\mathrm{ab}}$. In case there is no absorptive object in path b, the state of the photon after $N$ cycles is $\ket{10}_{\mathrm{ab}}$.
    }
    \label{fig:IFM}
\end{figure}

The concept of the IFM was first introduced by Dicke \cite{D:81:AJP}, and extended by Elitzur $et\text{ }al$ \cite{VE:93:FOP}. The idea is to ascertain the presence or absence of the absorptive object in the interferometer without interrogating it. 
We consider an array of $N$ unbalanced beam splitters (BS) \cite{KWHZK:95:PRL}  as shown in Fig.~\ref{fig:IFM}. Here, we represent the lower path as $\mathrm{a}$ and the upper path as $\mathrm{b}$. The BS in Fig.~\ref{fig:IFM} works as follows:
\begin{equation}
\begin{aligned}
\ket{10}_{\mathrm{ab}}\rightarrow\cos\theta_N\ket{10}_{\mathrm{ab}}+\sin\theta_N\ket{01}_{\mathrm{ab}},\\
\ket{01}_{\mathrm{ab}}\rightarrow\cos\theta_N\ket{01}_{\mathrm{ab}}-\sin\theta_N\ket{10}_{\mathrm{ab}},
\end{aligned}
\end{equation}
where $\theta_N=\pi/\left(2N\right)$, $\ket{10}_{\mathrm{ab}}$ denotes that the photon is in path $\mathrm{a}$, and $\ket{01}_{\mathrm{ab}}$ denotes that the photon is in path $\mathrm{b}$.

We consider that the initial state is $\ket{10}_{\mathrm{ab}}$. If there is no absorptive object in path b, the state of the photon after $n\left(<N\right)$ cycles is given by
\begin{equation}
\ket{10}_{\mathrm{ab}}\rightarrow\cos \left(n\theta_{N}\right)\ket{10}_{\mathrm{ab}}+\sin\left( n\theta_{N}\right)\ket{01}_{\mathrm{ab}}.
\label{eq:2}
\end{equation}
This photon ends up in the state $\ket{01}_{\mathrm{ab}}=\cos\left(N\theta_N\right)\ket{10}_{\mathrm{ab}}+\sin \left(N\theta_N\right)\ket{01}_{\mathrm{ab}}$ after $N$ cycles. In case the absorptive object blocks path b of the photon, the state of the photon after $n\left(<N\right)$ cycles is given by
\begin{equation}
\ket{10}_{\mathrm{ab}}\rightarrow \cos^{\left(n-1\right)}\theta_{N}\left(\cos\theta_{N}\ket{10}_{\mathrm{ab}}+\sin\theta_{N}\ket{01}_{\mathrm{ab}}\right).
\label{eq:3}
\end{equation}
After $N$ cycles, the existing photon ends up in the state $\ket{10}_{\mathrm{ab}}$. Hence, we can conclude that if the path of the photon is blocked by an absorptive object, the existing photon is in state $\ket{10}_{\mathrm{ab}}$ with probability $\cos^{2N}\theta_N$. If there is no absorptive object, photon ends up in the state $\ket{01}_{\mathrm{ab}}$ with probability one. 

Quantum superdense coding increases the classical capacity of the quantum channel by utilizing the preshared entanglement. In general, remote parties can transmit maximum 1 bit/qubit of classical information under the ideal channel conditions. In the presence of preshared entanglement between Alice (sender) and Bob (receiver), the quantum superdense coding doubles the data rate to 2 bits/qubit. Without loss of generality, we assume that Alice and Bob shared the bipartite maximally entangled state where Alice wants to send a two-bit classical message to Bob using quantum superdense coding. Let us define the throughput efficiency $R$ in bits/qubit for superdense coding by the average number of successfully delievered classical information bits per qubit transmission. Then, for the uniform distribution of classical bits $\lbrace 00,01,10,11\rbrace$, the throughput efficiency $R$ [bits/qubit] for quantum superdense coding with the IFM Bell-state analyzer  \cite{A:04:PRA} is given by (see~\hyperref[sec: Methods]{Methods})
\begin{align}
R=\left(\cos^{2N}\theta_N+1\right)\left(1-\dfrac{1}{2}\sin^2\theta_N\right)^N.\label{eq:4}
\end{align}

Later, the QZ CNOT gate using an ancillary photon has been presented where both the target bit and the control bit were considered as the quantum absorptive object \cite{HM:08:PRA}. To discriminate between the four orthonormal Bell states, the ancillary photon needs not to be discarded. The probability $P_{\text{QZ}}$ that the ancillary photon is not absorbed by the entangled particle(s) and throughput efficiency $R$ [bits/qubit] for quantum superdense coding with this QZ Bell-state analyzer are respectively given by (see~\hyperref[sec: Methods]{Methods})
\begin{align}\label{eq:5}
P_{\text{QZ}}&=\left(1-\dfrac{3}{4}\sin^2\phi_N\right)^N,
\\
\label{eq:6}
R&=2 P_{\text{QZ}},
\end{align}
where $\phi_N=\pi/N$. Note that the QZ Bell-state analyzer sacrifices the throughput efficiency for quantum superdense coding in order to improve the resource efficiency. In this paper, we enhance the throughput efficiency (or both the resource efficiency and the throughput efficiency) for quantum superdense coding by using the DQZ gate instead of using the QZ gate (or multiple IFM gates) for the Bell-state analyzer.

\section*{Results}
\label{sec: Results}
\begin{figure}[t!]
 \centering
\includegraphics[width=0.7\textwidth]{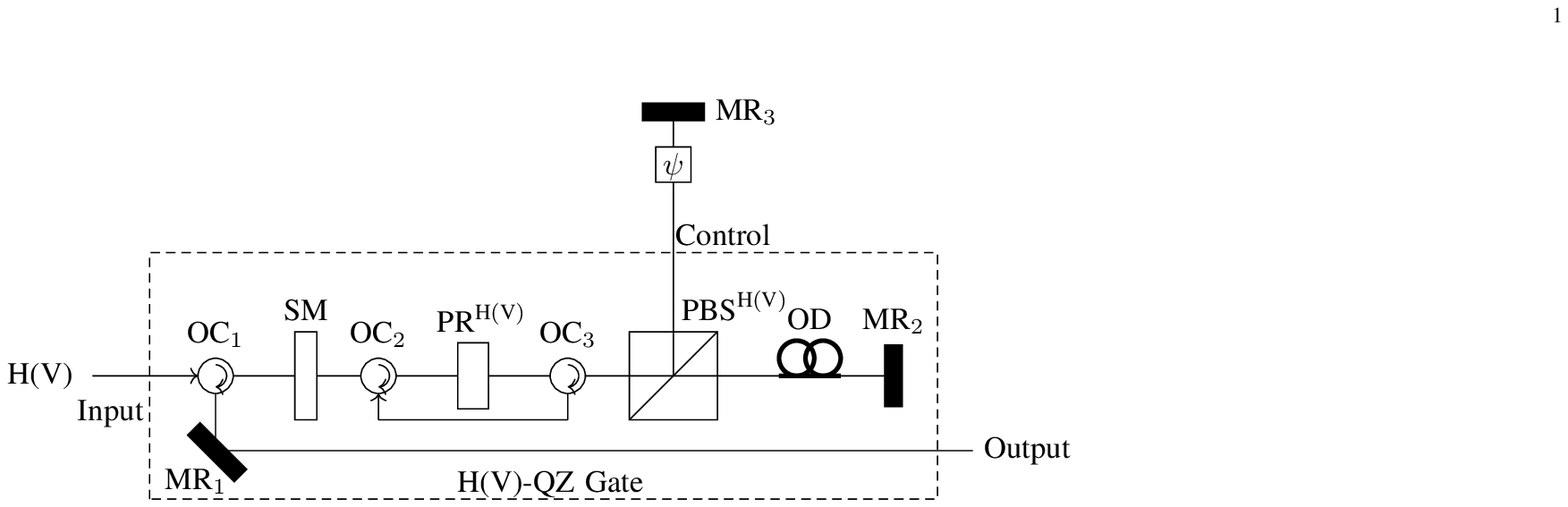}
    \caption{
   \textbf{Michelson version of the $\mathrm{H}\left(\mathrm{V}\right)$-QZ gate.} Here SM is a switchable mirror operated by external means to control the motion of the photon, PR is a polarization rotator, PBS is a polarizing beam splitter which isolates the H and V components of the photon. The combined action of PR and PBS achieves the function of BS in Fig.~\ref{fig:IFM}. MR is a mirror. OD speaks to an optical deferral, OC is an optical circulator and $\psi$ denotes the state of the absorptive object. The absorption probability of the photon is asymptotically zero as $N$ approaches to $\infty$. The overall action of the $\mathrm{H}\left(\mathrm{V}\right)$-QZ gate is given in Table~\ref{tab:I-QZE}.
    }
    \label{fig:QZE}
\end{figure}
\begin{table}[t!]
\centering
\setlength{\tabcolsep}{8pt}
{\renewcommand{\arraystretch}{1.5}
\caption{H(V)-QZ gate as $N \rightarrow \infty$ where $\ket{-}_{\mathrm{p}}$ shows that the photon is discarded.}\label{tab:I-QZE}
\vspace{0.2cm}
\begin{tabular}{ p{2.5cm}| p{2.5cm}}
\hline
 Input & Output  \\ [0.5ex]
\hline
\hline 
$\ket{\text{pass}}_{\psi}\ket{\mathrm{H}\left(\mathrm{V}\right)}_{\mathrm{p}}$ & $\ket{\text{pass}}_{\psi}\ket{\mathrm{V}\left(\mathrm{H}\right)}_{\mathrm{p}}$\\
$\ket{\text{pass}}_{\psi}\ket{\mathrm{V}\left(\mathrm{H}\right)}_{\mathrm{p}}$ & $\ket{\text{pass}}_{\psi}\ket{\mathrm{H}\left(\mathrm{V}\right)}_{\mathrm{p}}$\\
$\ket{\text{block}}_{\psi}\ket{\mathrm{H}\left(\mathrm{V}\right)}_{\mathrm{p}}$ & $\ket{\text{block}}_{\psi}\ket{\mathrm{H}\left(\mathrm{V}\right)}_{\mathrm{p}}$\\
$\ket{\text{block}}_{\psi}\ket{\mathrm{V}\left(\mathrm{H}\right)}_{\mathrm{p}}$ & $\ket{\text{block}}_{\psi}\ket{-}_{\mathrm{p}}$\\[0.5ex]
\hline
\end{tabular}}
\end{table}

\subsection*{DQZ Gate}

In this section, we consider the Michelson version of the QZ gate \cite{KWMNWWZ:99:PRL}. Fig.~\ref{fig:QZE} shows in principle how to perform an IFM with the QZ gate, where $\mathrm{H}\left(\mathrm{V}\right)$ refers to horizontal $\left(\text{vertical}\right)$ polarization. The operation performed by BS in IFM setup is achieved by performing the polarization rotator (PR) followed by the polarizing beam splitter (PBS) as shown in Fig.~\ref{fig:QZE}.   The action of $\text{PR}^{\mathrm{H}\left(\mathrm{V}\right)}$ on the input photon is
\begin{align}
\text{PR}^{\mathrm{H}}
&\begin{cases}
\ket{\mathrm{H}}_{\mathrm{p}}\rightarrow\cos\theta_{N}\ket{\mathrm{H}}_{\mathrm{p}}+\sin\theta_{N}\ket{\mathrm{V}}_{\mathrm{p}},\\
\ket{\mathrm{V}}_{\mathrm{p}}\rightarrow\cos\theta_{N}\ket{\mathrm{V}}_{\mathrm{p}}-\sin\theta_{N}\ket{\mathrm{H}}_{\mathrm{p}},
\end{cases}
\\
\text{PR}^{\mathrm{V}}
&\begin{cases}
\ket{\mathrm{V}}_{\mathrm{p}}\rightarrow\cos\theta_{N}\ket{\mathrm{V}}_{\mathrm{p}}+\sin\theta_{N}\ket{\mathrm{H}}_{\mathrm{p}},\\
\ket{\mathrm{H}}_{\mathrm{p}}\rightarrow\cos\theta_{N}\ket{\mathrm{H}}_{\mathrm{p}}-\sin\theta_{N}\ket{\mathrm{V}}_{\mathrm{p}}.
\end{cases}
\end{align}
where $\ket{\mathrm{H}}_{\mathrm{p}}=\ket{1}_{\mathrm{p}}$ and $\ket{\mathrm{V}}_{\mathrm{p}}=\ket{0}_{\mathrm{p}}$; and the subscript p denotes the photon. The Michelson version of the H(V)-QZ gate takes an $\mathrm{H}\left(\mathrm{V}\right)$ polarized photon as input, with $\text{PBS}^{\mathrm{H}\left(\mathrm{V}\right)}$ which allows the H (V) component of the photon to pass and reflects the V (H) component of the photon as shown in Fig.~\ref{fig:QZE}. The basic idea behind the IFM and QZ gates \cite{KWHZK:95:PRL,KWMNWWZ:99:PRL} was to ascertain the classical behavior of the absorptive object, i.e., infer the presence or absence of the bomb in the interferometer without interacting with it. This idea has been extended to the quantum absorptive object such as the positron \cite{A:04:PRA} and atom \cite{HM:08:PRA} so that it can take the superposition of pass and block. Let us consider how the QZ gate in Fig.~\ref{fig:QZE} works if the absorptive object is in the superposition of pass and block. The H(V)-QZ gate takes H (V) polarized photon as input. From Table~\ref{tab:I-QZE}, for the superposition corresponding to $\ket{\text{pass}}_{\psi}$, the output photon is $\mathrm{V}\left(\mathrm{H}\right)$ polarized, while for the superposition corresponding to $\ket{\text{block}}_{\psi}$, the existing photon is $\mathrm{H}\left(\mathrm{V}\right)$ polarized.

\begin{figure}[t!]
 \centering
\includegraphics[width=0.6\textwidth]{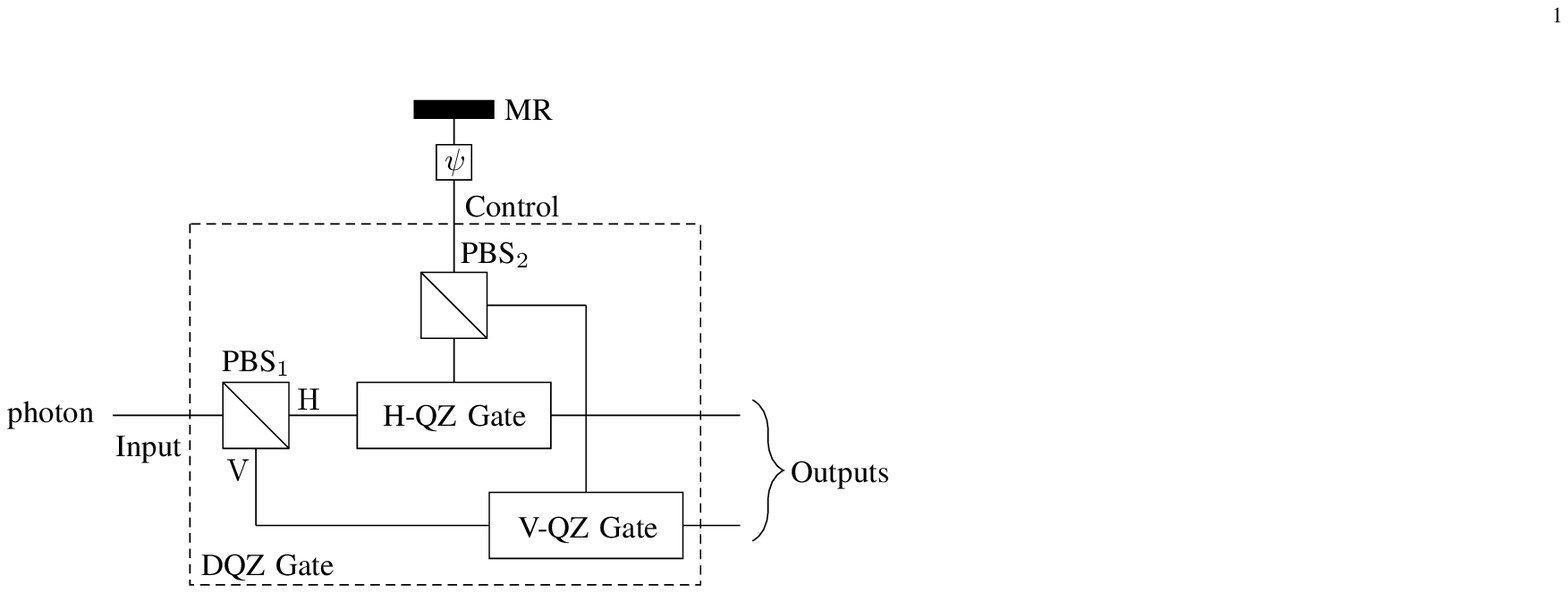}
    \caption{
  \textbf{DQZ gate.}
  The protocol starts by throwing the photon from the left towards $\mathrm{PBS}_1$ to separate each polarized component of the photon and feed into the corresponding QZ gate. The $\mathrm{PBS}_2$ recombines the $\mathrm{H}$ and $\mathrm{V}$ components for the simultaneous interaction with the quantum absorptive object. As $\ket{\mathrm{H}}_{\mathrm{p}}$ and $\ket{\mathrm{V}}_{\mathrm{p}}$ are orthogonal to each other, the simultaneous interaction with the quantum absorptive object does not effect the counterparts. The general activity of the DQZ entryway carries on like a CNOT door where the control bit is the quantum absorptive object and the photon goes about as a target bit.
    }
    \label{fig:d-QZE}
\end{figure}

If the photon is in the superposition state $\alpha\ket{\mathrm{H}}_{\mathrm{p}}+\beta\ket{\mathrm{V}}_{\mathrm{p}}$ in the presence of the absorptive object,  the likelihood that the photon is not disposed of in the H(V)-QZ gate is $|\alpha|^2\left(|\beta|^2\right)$ under the asymptotic limit of $N$. We first pass the photon through the $\text{PBS}_1$  to improve the efficiency up to $100\%$ as shown in Fig.~\ref{fig:d-QZE}. The PBS separates the each polarized component of the photon and feed into  the corresponding QZ gate. Then, the composite state of the photon and the quantum absorptive object is given by
\begin{align}
\left(\lambda\ket{\text{pass}}_{\psi}+\mu\ket{\text{block}}_{\psi}\right)
\left(\alpha\ket{\mathrm{H}}_{\mathrm{p}}+\beta\ket{\mathrm{V}}_{\mathrm{p}}\right).
\end{align}
As $N \rightarrow \infty$, this combined state  after $N$ cycles becomes
\begin{equation}
\alpha\lambda\ket{\text{pass}}_{\psi}\ket{\mathrm{V}}_{\mathrm{p}}+\beta\lambda\ket{\text{pass}}_{\psi}\ket{\mathrm{H}}_{\mathrm{p}}+
\alpha\mu\ket{\text{block}}_{\psi}\ket{\mathrm{H}}_{\mathrm{p}}+\beta\mu\ket{\text{block}}_{\psi}\ket{\mathrm{V}}_{\mathrm{p}}.
\label{eq:9}
\end{equation}
Using equivalent binary states,~\eqref{eq:9} can be rewritten as
\begin{equation}
\alpha\left(\lambda\ket{1}_{\text{AO}}\ket{0}_{\mathrm{p}}+\mu\ket{0}_{\text{AO}}\ket{1}_{\mathrm{p}}\right)+
\beta\left(\lambda\ket{1}_{\text{AO}}\ket{1}_{\mathrm{p}}+\mu\ket{0}_{\text{AO}}\ket{0}_{\mathrm{p}}\right),
\end{equation}
where $\ket{0}_{\text{AO}}=\ket{\text{block}}_{\psi}$ and $\ket{1}_{\text{AO}}=\ket{\text{pass}}_{\psi}$, respectively. This is our DQZ gate. 
\subsection*{Bell-State Analysis}

\begin{figure*}[t!]
 \centering
\includegraphics[width=1\textwidth]{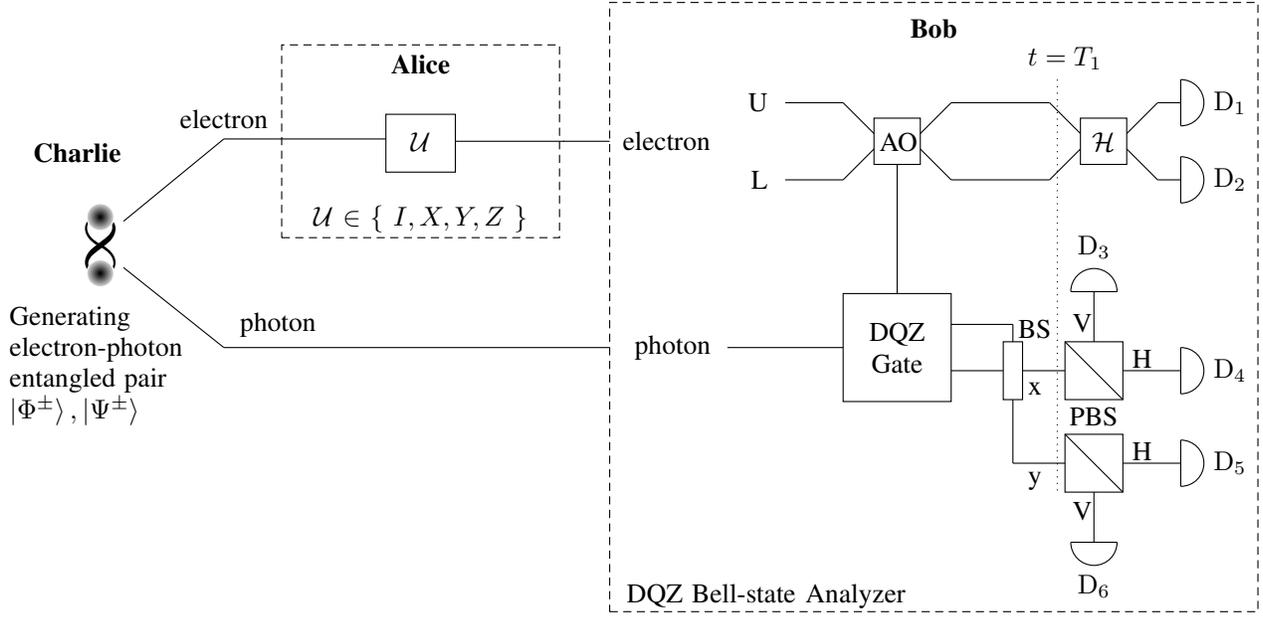}
    \caption{
  \textbf{DQZ superdense coding.} Here $X$, $Y$, and $Z$ represent Pauli operators $\sigma_x$, $\sigma_y$, and $\sigma_z$, respectively.  AO stands for an absorptive object. Initially, Alice and Bob receive a maximally entangled electron--photon pair. Alice performs one of the four unitary operators on her entangled particle to encode two bits of classical information. Alice sends her entangled particle (electron) to Bob who performs the complete Bell-sate analysis where the electron act as quantum absorptive object for the DQZ gate. The presence and absence of the quantum absorptive object (electron) are represented as $\ket{10}_{\mathrm{LU}}$ and $\ket{01}_{\mathrm{LU}}$, where the subscripts L and U denote the lower and upper paths, respectively. Bob starts by throwing his photon towards the DQZ gate. After the DQZ gate, Bob measures the polarization of the existing photon, which enables Bob to extract only one bit of classical information. To decode two bits of classical information, Bob performs the Hadamard gate $\Hb$ on the electron and measures the path of the electron. We estimate the classical message; and the initial Bell state corresponding to the detector clicks (see Table~\ref{tab:Measurement Result}). 
 }
    \label{fig:SDC}
\end{figure*}

To demonstrate the DQZ Bell-state analyzer, we consider that the composite state of the electron-photon pair is in one of the four Bell states where the electron is the quantum absorptive object. The Bell states are given by
\begin{align}
\ket{\Phi^{\pm}}&=\dfrac{1}{\sqrt{2}}\left(\ket{\text{block}}_{\mathrm{e}}\ket{\mathrm{V}}_{\mathrm{p}}\pm\ket{\text{pass}}_{\mathrm{e}}\ket{\mathrm{H}}_{\mathrm{p}}\right),\label{eq: Phi states}\\
\ket{\Psi^{\pm}}&=\dfrac{1}{\sqrt{2}}\left(\ket{\text{block}}_{\mathrm{e}}\ket{\mathrm{H}}_{\mathrm{p}}\pm\ket{\text{pass}}_{\mathrm{e}}\ket{\mathrm{V}}_{\mathrm{p}}\right),\label{eq: Psi states}
\end{align}
where $\ket{\text{pass}}_{\mathrm{e}}=\ket{1}_{\mathrm{e}}=\ket{01}_{\text{LU}}$, and $\ket{\text{block}}_{\mathrm{e}}=\ket{0}_{\mathrm{e}}=\ket{10}_{\text{LU}}$;  and the subscript $\mathrm{e}$ is the electron, L and U are the two paths in Fig.~\ref{fig:SDC} (DQZ Bell-state Analyzer), respectively. If the state of the electron is $\ket{10}_{\text{LU}}$, it represents the presence of the absorptive object (electron). In case the state of the electron is $\ket{01}_{\text{LU}}$, it represents the absence of the absorptive object. The protocol starts by throwing the photon from the left towards the DQZ gate as shown in Fig.~\ref{fig:SDC} (DQZ Bell-state Analyzer). At $t=T_1$, the composite state of the electron-photon pair for large $N$ is given as  
\begin{align}
\ket{\Phi^{\pm}}\rightarrow\dfrac{1}{\sqrt{2}}\left(\ket{\text{block}}_{\mathrm{e}}\pm\left(-1\right)^{\mathrm{m}}\ket{\text{pass}}_{\mathrm{e}}\right)\ket{\mathrm{V}}_{\mathrm{p}},
\label{eq: Phi seperable}\\
\ket{\Psi^{\pm}}\rightarrow\dfrac{1}{\sqrt{2}}\left(\ket{\text{block}}_{\mathrm{e}}\pm\left(-1\right)^{\mathrm{m}}\ket{\text{pass}}_{\mathrm{e}}\right)\ket{\mathrm{H}}_{\mathrm{p}}.
\label{eq: Psi seperable}
\end{align} 
where $\mathrm{m}=0~\text{or}~1$ if the photon is in path x or y. After the $50:50$  BS, the polarization of the  photon distinguishes between $\ket{\Phi^{\pm}}$ and $\ket{\Psi^{\pm}}$. To estimate the initial Bell state, we perform the Hadamard gate on the electron and measure the path of the electron, which transform the state of the electron as follows:
\begin{align}
\dfrac{1}{\sqrt{2}}\left(\ket{\text{block}}_{\mathrm{e}}+\ket{\text{pass}}_{\mathrm{e}}\right)\rightarrow\ket{\text{block}}_{\mathrm{e}}&=\ket{10}_{\mathrm{LU}},\\
\dfrac{1}{\sqrt{2}}\left(\ket{\text{block}}_{\mathrm{e}}-\ket{\text{pass}}_{\mathrm{e}}\right)\rightarrow\ket{\text{pass}}_{\mathrm{e}}&=\ket{01}_{\mathrm{LU}},
\end{align}
where $\ket{10}_{\mathrm{LU}}$ shows that the electron is in path $\mathrm{L}$ and vice versa. We gauge the underlying state with assurance relating to the detector clicks (see Table~\ref{tab:Measurement Result}). The complete Bell-state analysis can also be performed by two remote parties in a semi-counterfactual way (see~\hyperref[sec: Discussion]{Discussion}).

\begin{figure}[t]
 \centering
\includegraphics[width=0.55\textwidth]{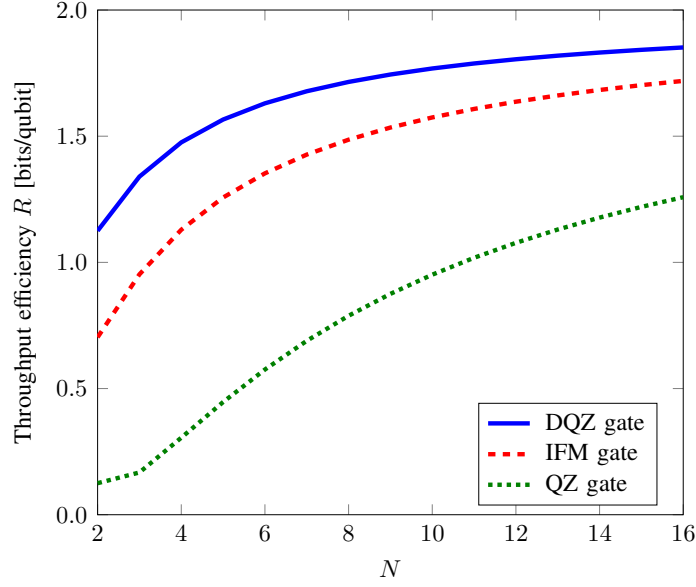}
    \caption{
   Throughput efficiency $R$ [bits/qubit] for quantum superdense coding as a function of the number of cycles $N$ with the advantage of complete Bell-state analysis based on the DQZ gate, IFM gate \cite{A:04:PRA}, and QZ gate \cite{HM:08:PRA}, respectively. }
    \label{fig:plot}
\end{figure}
\begin{table}[t!]
\centering
\setlength{\tabcolsep}{8pt}
{\renewcommand{\arraystretch}{1.5}
\caption{DQZ superdense coding as $N \rightarrow \infty$.}\label{tab:Measurement Result}
\vspace{0.2cm}
\begin{tabular}{  p{1.7cm}| p{1.7cm} | p{1.7cm}}
\hline
\multicolumn{3}{c}{After Hadamard Gate} \\
\hline
  Detector Clicks & Estimated Initial State & Classical Message\\ [0.5ex]
\hline
\hline 
 $\mathrm{D}_{1\left(2\right)}\mathrm{D}_{3\left(6\right)}$ & $\ket{\Phi^-}$ & 10 \\
 $\mathrm{D}_{2\left(1\right)}\mathrm{D}_{3\left(6\right)}$ & $\ket{\Phi^+}$ & 00\\
 $\mathrm{D}_{1\left(2\right)}\mathrm{D}_{4\left(5\right)}$ & $\ket{\Psi^-}$  & 11\\
 $\mathrm{D}_{2\left(1\right)}\mathrm{D}_{4\left(5\right)}$ & $\ket{\Psi^+}$ & 01 \\[0.5ex]
\hline
\end{tabular}}
\end{table}

\subsection*{Quantum Superdense Coding}

We consider that a third party (Charlie) generates an entangled electron--photon pair $\left(\ket{\Phi^+}\right)$ and transmit the electron to Alice $\left(\text{sender}\right)$ and  the photon to Bob $\left(\text{receiver}\right)$, respectively. Alice applies unitary operation 
$\Ub\in\left\{I,X,Y,Z\right\}$ corresponding to classical bits she wants to transmit as follows: 
\begin{equation}
\begin{aligned}
00&\rightarrow\left(I_{\mathrm{A}}\otimes I_{\mathrm{B}}\right)\ket{\Phi^+}=\ket{\Phi^+},\\
01&\rightarrow\left(X_{\mathrm{A}}\otimes I_\mathrm{B}\right)\ket{\Phi^+}=\ket{\Psi^+},\\
10&\rightarrow\left(Z_{\mathrm{A}}\otimes I_\mathrm{B}\right)\ket{\Phi^+}=\ket{\Phi^-},\\
11&\rightarrow\left(Y_{\mathrm{A}}\otimes I_\mathrm{B}\right)\ket{\Phi^+}=\ket{\Psi^-},
\end{aligned}
\end{equation} 
where $I$ is the identity operator; $X$, $Y$, and $Z$ represent Pauli operators $\sigma_x$, $\sigma_y$, and $\sigma_z$, respectively; and the subscripts $\mathrm{A}$ and $\mathrm{B}$ indicate the operators acting on Alice's and Bob's sides.  Consequently, the initial entangled pair is changed to one of the four orthonormal Bell states. Then, Alice sends her entangled particle to Bob who performs the DQZ Bell-state analysis to decode the classical message as shown in Fig.~\ref{fig:SDC}. After the Hadamard gate, Bob measures the state of the electron and the polarization of the exciting photon to decode the classical information encoded in Bell states with likelihood one. Table~\ref{tab:Measurement Result} demonstrate the decoded classical messages corresponding to the detectors click. As $N \rightarrow \infty$, the throughput efficiency  for quantum superdense coding approaches to 2 bits/qubit. For a finite value of $N$, there is a nonzero probability that the photon is absorbed by the electron.  This absorbing probability in each cycle of the DQZ gate is equal to
\begin{align}
\begin{cases}
0,~&\text{if electron is in }\ket{\text{pass}}_{\mathrm{e}},\\
\sin^2\theta_N,~&\text{if electron is in }\ket{\text{block}}_{\mathrm{e}}.
\end{cases}
\label{eq: probabilites}
\end{align}
Since each Bell state is in equal superposition of $\ket{\text{pass}}_{\mathrm{e}}\ket{\mathrm{H}\left(\mathrm{V}\right)}_{\mathrm{p}}$ and $\ket{\text{block}}_{\mathrm{e}}\ket{\mathrm{V}\left(\mathrm{H}\right)}_{\mathrm{p}}$, as seen from~\eqref{eq: Phi states} and~\eqref{eq: Psi states}, the probability $P$ that the photon is not discarded in each cycle for any input Bell state is 
\begin{align}
P=1-\dfrac{1}{2}\sin^2\theta_N.
\end{align}

Let $\rho_0^{\pm}=\ket{\Psi^{\pm}}\bra{\Psi^{\pm}}$,  $\rho_1^{\pm}=\ket{\Phi^{\pm}}\bra{\Phi^{\pm}}$, and 
\begin{align}
K_i&=\begin{pmatrix}
1 & 0 & 0 & 0\\
0 & 1 & 0 & 0 \\
0 & 0 & \cos\theta_N & \left(-1\right)^{i+1}\sin\theta_N\\
0 & 0 & \left(-1\right)^{i}\sin\theta_N & \cos\theta_N
\end{pmatrix}.
\end{align}
Then, after the first cycle, we have the state transformations
\begin{align}
\rho_i^{\pm}\rightarrow PK_i\rho_i^\pm K_i^{\textit{T}}+\left(1-P\right)\ket{0}_{\mathrm{e}}\hspace{-0.07cm}\bra{0}\otimes\ket{i}_{\mathrm{p}}\hspace{-0.07cm}\bra{i} \label{eq:19},
\end{align}
for $i=0,1$, where $\left(\cdot\right)^{\textit{T}}$ denotes the matrix transpose. The first term in~\eqref{eq:19} shows the state of the composite system if the photon is not discarded.
In case the photon is absorbed by the electron, the state of the composite system collapses to the second term in~\eqref{eq:19}. 
Hence, after $N$ cycles, the states transform as
\begin{align}
\rho_i^\pm \rightarrow P^NK_i^N \rho_i^\pm  \left(K_i^{\textit{T}}\right)^N+\left(1-P^N\right)\ket{0}_{\mathrm{e}}\hspace{-0.07cm}\bra{0}\otimes\ket{i}_{\mathrm{p}}\hspace{-0.07cm}\bra{i}.
\end{align}
Since each Bell state carries a two-bit classical message, the throughput efficiency $R$ [bits/qubit] is given by
\begin{align}
R=
2 \left(1-\dfrac{1}{2}\sin^2\theta_N\right)^N.
\end{align}
\begin{table}[t!]
\centering
\setlength{\tabcolsep}{8pt}
{\renewcommand{\arraystretch}{1.5}
\caption{Comparison of quantum superdense coding using the QZ, IFM and DQZ gates.}\label{tab:comparison}
\vspace{0.2cm}
\begin{tabular}{ p{7cm} p{1.5cm} p{1.5cm} p{1.5cm}}
\hline
 Metric & QZ Gate & IFM Gate & DQZ Gate\\ [0.5ex]
\hline
\hline 
$N$ to attain $R=1.8$ [bits/qubit] ($90\%$ efficiency) & 71 & 24 & 12 \\
Number of beamsplitters 	   & $N$ & $4N$ & $2N$ \\
Necessity of ancillary particle				   & Yes & No & No\\
Completeness of Bell-state analysis   & Yes & Yes & Yes \\
Feasibility of Bell-state analysis for remote parties & No & Yes & Yes \\[0.5ex]
\hline
\end{tabular}}
\end{table}

Fig.~\ref{fig:plot} shows the throughput efficiency for quantum superdense coding as a function of $N$ with the advantage of complete Bell-state analysis based on the DQZ gate, IFM gate \cite{A:04:PRA}, and QZ gate \cite{HM:08:PRA}, respectively. We observe that the DQZ superdense coding achieves higher throughput efficiency as compared to using the IFM and QZ gates for any value of $N$. Recently, the largest throughput efficiency $R=1.665$ [bits/qubit] has been achieved experimentally \cite{WRT:17:PRL} by quantum superdense coding with the hyperentanglement-assisted Bell-state analyzer where the time is an ancillary degree of freedom. The DQZ superdense coding attains the throughput efficiency larger than this experimental throughput when $N\geq 7$. For example, $R=1.678$ [bits/qubit] when $N=7$.

\section*{Discussion}
\label{sec: Discussion}

In this paper, we presented the new scheme for quantum superdense coding based on the DQZ Bell-state analyzer. It has been shown that the DQZ Bell-state analyzer estimates the initial Bell state with probability one and  enhances the throughout efficiency for quantum superdense coding under limited resources (see Table~\ref{tab:comparison}).
Unlike the QZ Bell-state analyzer \cite{HM:08:PRA}, our scheme also enables remote parties to discrimintate between the orthonormal Bell states in a semi-counterfactual way. The QZ gate is counterfactual only for one classical bit ($\ket{\mathrm{block}}_{\mathrm{e}}$) which makes it semi-counterfactual or partially-counterfactual \cite{SLAZ:13:PRL}. 

\begin{figure}[t]
 \centering
\includegraphics[width=0.5\textwidth]{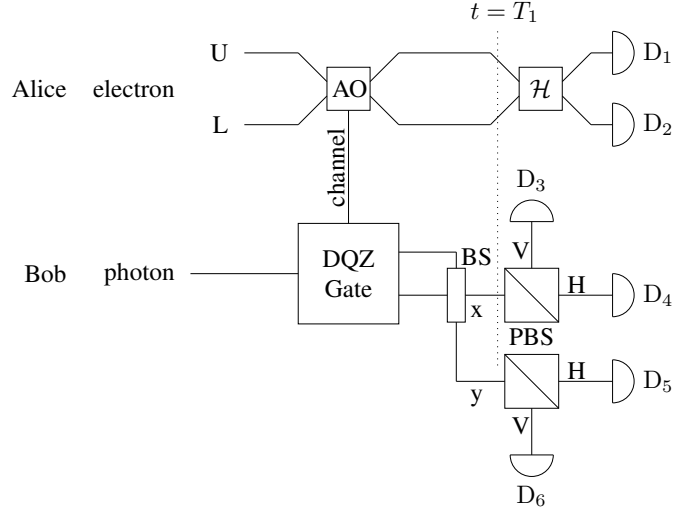}
    \caption{
   \textbf{Semi-counterfactual Bell-state analysis.} At first, Alice's and Bob's particles (electron and photon) are entangled. Bob starts the protocol by tossing  his entangled particle towards the DQZ gate. After BS, the polarization of the current photon decides either the underlying state $\ket{\Phi^{\pm}}$ or $\ket{\Psi^{\pm}}$. To discriminate between the four orthonormal Bell states, Alice performs the Hadamard gate $\Hb$ on the electron and sends her measurement result to Bob. As in Table~\ref{tab:Measurement Result}, Bob estimates the initial Bell state with certainty corresponding to the detector clicks. }
    \label{fig:BBM}
\end{figure}

Fig.~\ref{fig:BBM}. shows the schematic for distinguishability of four orthonormal Bell states in a semi-counterfactual way. The two remote parties, Alice and Bob, have an electron-photon pair in maximally entangled state. In the DQZ gate, if the composite state of the electron and photon is $\ket{\text{pass}}_{\mathrm{e}}\ket{\mathrm{H}\left(\mathrm{V}\right)}_{\mathrm{p}}$, there is a to-and-fro motion of a physical particle (photon) in the quantum channel, which makes our scheme semi-counterfactual. In case the electron-photon pair is in the state $\ket{\text{block}}_{\mathrm{e}}\ket{\mathrm{H}\left(\mathrm{V}\right)}_{\mathrm{p}}$, in the event that the photon is in the quantum channel, it will be consumed by the electron. After $N$ cycles, the likelihood that the photon is not discarded for any input Bell state is $P^N$.
After the DQZ gate, Alice and Bob perform a measurement on their respective particles. The polarization of the photon enables Bob to distinguish between $\ket{\Phi^{\pm}}$ and $\ket{\Psi^{\pm}}$. To perform the complete Bell-state analysis, Alice sends her measurement result to Bob using counterfactual communication\cite{SLAZ:13:PRL}. As in Table~\ref{tab:Measurement Result}, Bob estimates the initial state with certainty based on the measurement results of the electron and photon. To discriminate between the four orthonormal Bell states in a fully counterfactual way, Alice and Bob use the nested version of QZ gate at the cost of a large number of cycles requried for high probability of success \cite{ZJS:18:SR}.
\section*{Methods}
\label{sec: Methods}
\subsection*{Superdense Coding with IFM Bell-State Analyzer}

\begin{figure}[t]
 \centering
\includegraphics[width=0.95\textwidth]{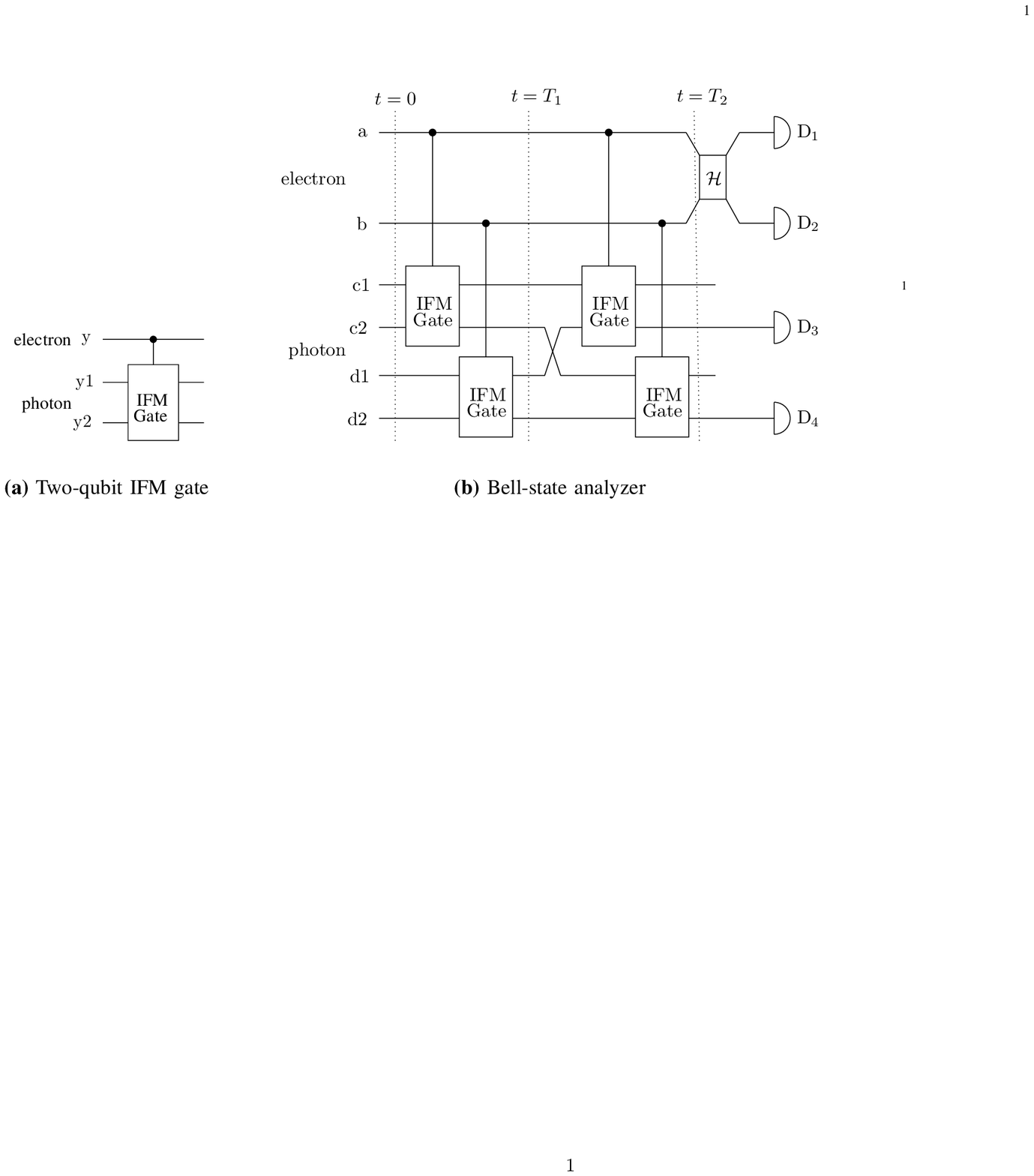}
    \caption{
   IFM Bell-state analyzer. Initially, the photon is in equal superposition of c2 and d2. At $t=T_2$, if the photon ends up at the detector D$_3$, the initial state is $\ket{\Psi^{\pm}}$. In case the photon ends up at the detector D$_4$, the initial state is $\ket{\Phi^{\pm}}$. Bob applies the Hadamard gate on the state of the electron and estimates the initial state corresponding to the detectors click.   }
    \label{fig:fig7}
\end{figure}

The two-qubit IFM gate changes the trajectory of the photon according to whether or not the absorptive object is in path $y$ as shown in Fig.~\ref{fig:fig7}\textbf{(a)}. If the initial state of the photon is $\ket{0}_{\mathrm{y1}}\ket{1}_{\mathrm{y2}}$, in the nearness or nonappearance of the absorptive object, the photon will end up in the state $\ket{0}_{\mathrm{y1}}\ket{1}_{\mathrm{y2}}$ or $\ket{1}_{\mathrm{y1}}\ket{0}_{\mathrm{y2}}$.  To discriminate between the four orthonornal Bell states \cite{A:04:PRA}, we consider an entangled pair of the electron and photon. At $t=0$, the Bell states are written as: 
\begin{align}
\ket{\Phi^{\pm}}
&=\dfrac{1}{\sqrt{2}}\bigl(\underbrace{\ket{0}_{\mathrm{e}}\ket{0}_{\mathrm{p}}\pm\ket{1}_{\mathrm{e}}\ket{1}_{\mathrm{p}}}_{\text{block}}\bigr),\\
\ket{\Psi^{\pm}}
&=\dfrac{1}{\sqrt{2}}\bigl(\underbrace{\ket{0}_{\mathrm{e}}\ket{1}_{\mathrm{p}}\pm\ket{1}_{\mathrm{e}}\ket{0}_{\mathrm{p}}}_{\text{pass}}\bigr),
\label{eq: Bell-states}
\end{align} 
where $\ket{0}_{\mathrm{e}}=\ket{0}_{\mathrm{a}}\ket{1}_{\mathrm{b}}$, $\ket{1}_{\mathrm{e}}=\ket{1}_{\mathrm{a}}\ket{0}_{\mathrm{b}}$, $\ket{0}_{\mathrm{p}}=\ket{0}_{\mathrm{c1}}\ket{0}_{\mathrm{c2}}\ket{0}_{\mathrm{d1}}\ket{1}_{\mathrm{d2}}$ and $\ket{1}_{\mathrm{p}}=\ket{0}_{\mathrm{c1}}\ket{1}_{\mathrm{c2}}\ket{0}_{\mathrm{d1}}\ket{0}_{\mathrm{d2}}$; and the subscripts e, p, a, b, $\mathrm{c1}$ c2, $\mathrm{d1}$ and d2 denote the electron, photon and the paths in Fig.~\ref{fig:fig7}\textbf{(b)}. Here, $\ket{0}_{\mathrm{z}}$ and $\ket{1}_{\mathrm{z}}$  show the absence and presence of the particle in the respective path, where $\mathrm{z}\in\lbrace \mathrm{a,b, }\mathrm{c1}\mathrm{,c2,}\mathrm{d1} \mathrm{,d2}\rbrace$. At $t=T_1$, the average probability $P_{\text{avg}}$ that the photon is not absorbed by the electron corresponding to each Bell state is given as:
\begin{equation}
\begin{aligned}
P_{\text{avg}}=\dfrac{1}{2}\left(\cos^{2N}\theta_N+1\right),
\label{eq: average probability}
\end{aligned} 
\end{equation}
and the Bell states transform as:
\begin{align}
\ket{\Phi^{\pm}}
&
\rightarrow\dfrac{1}{\sqrt{2}}\bigl(\underbrace{\ket{0}_{\mathrm{a}}\ket{1}_{\mathrm{b}}\ket{0}_{\mathrm{c1}}\ket{0}_{\mathrm{c2}}\ket{0}_{\mathrm{d1}}\ket{1}_{\mathrm{d2}}}_{\text{block}}\pm\underbrace{\ket{1}_{\mathrm{a}}\ket{0}_{\mathrm{b}}\ket{0}_{\mathrm{c1}}\ket{1}_{\mathrm{c2}}\ket{0}_{\mathrm{d1}}\ket{0}_{\mathrm{d2}}}_{\text{pass}}\bigr),\\
\ket{\Psi^{\pm}}
&
\rightarrow\dfrac{1}{\sqrt{2}}\bigl(\underbrace{\ket{0}_{\mathrm{a}}\ket{1}_{\mathrm{b}}\ket{1}_{\mathrm{c1}}\ket{0}_{\mathrm{c2}}\ket{0}_{\mathrm{d1}}\ket{0}_{\mathrm{d2}}}_{\text{pass}}\pm\underbrace{\ket{1}_{\mathrm{a}}\ket{0}_{\mathrm{b}}\ket{0}_{\mathrm{c1}}\ket{0}_{\mathrm{c2}}\ket{1}_{\mathrm{d1}}\ket{0}_{\mathrm{d2}}}_{\text{block}}\bigr).
\label{eq: Bell-states T_1}
\end{align}
Since each Bell state in equal superposition of pass and block carries two-bit classical message, using~\eqref{eq: probabilites} and~\eqref{eq: average probability}, we obtain the throughput efficiency $R$ [bits/qubit] as:
\begin{align}
R=2P_{\text{avg}}\left(1-\dfrac{1}{2}\sin^2\theta_N\right)^N.
\end{align}
\subsection*{Superdense Coding with QZ Bell-State Analyzer}
\begin{figure}[t!]
 \centering
\includegraphics[width=0.3\textwidth]{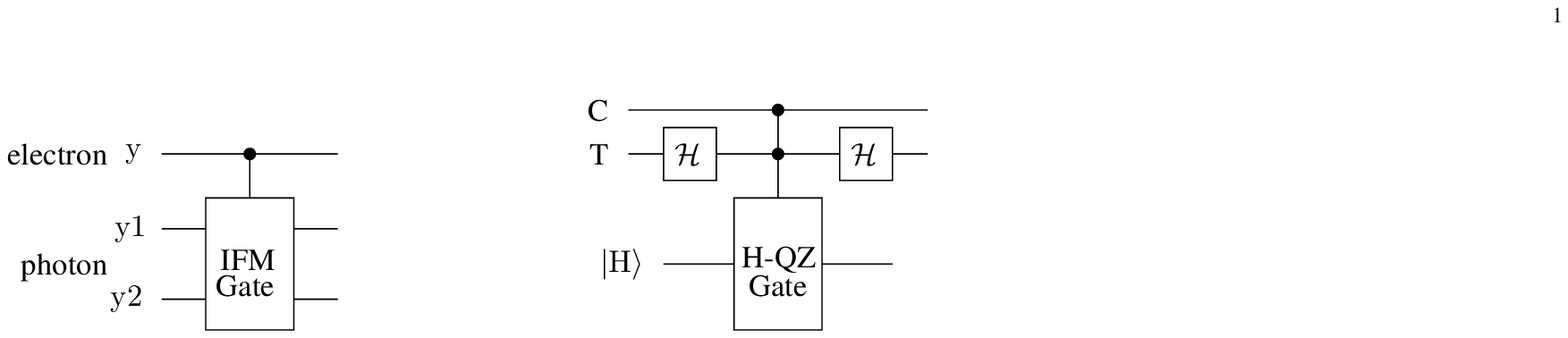}
    \caption{
   QZ Bell-state analyzer. Here C stands for the control bit and T is the target bit. The protocol starts by sending the H polarized ancillary photon towards the H-QZ gate which undergoes $\pi$ rotation only for $\ket{1}_{\mathrm{C}}\ket{1}_{\mathrm{T}}$. To discriminate between the four orthonormal Bell states, the ancillary photon needs not to be discarded.}
    \label{fig:QZ_BBM}
\end{figure}

In general, the QZ gate \cite{KWMNWWZ:99:PRL} rotates the input state of the photon by an angle $\theta$ based on the state of the absorptive object. Table~\ref{tab:I-QZE} demonstrates the general activity of the QZ gate for $\theta=\pi/2$. Based on this basic idea, the QZ Bell-state analyzer \cite{HM:08:PRA} has been presented by considering both the entangled particles as quantum absorptive objects and $\theta=\pi$ as shown in Fig.~\ref{fig:QZ_BBM}. One entangled particle is treated as a control bit (C) and the other serves as a target bit (T). 

To demonstrate the complete Bell-state analysis, we consider that the initial Bell-state is $\ket{\Phi^{+}}$ and the ancillary photon is H polarized. The protocol starts by applying the Hadamard transformation on the target bit as shown in Fig.~\ref{fig:QZ_BBM}. The input state $\ket{\Psi_{\text{in}}}$ of the system transforms as
\begin{align}
\ket{\Psi_\text{in}}
&
=\ket{\mathrm{H}}\otimes\dfrac{1}{\sqrt{2}}\left(\ket{0}_{\mathrm{C}}\ket{0}_{\mathrm{T}}+\ket{1}_{\mathrm{C}}\ket{1}_{\mathrm{T}}\right) \nonumber \\
&
\xrightarrow[]{\Hb}\ket{\mathrm{H}}\otimes\dfrac{1}{2}\bigl(\underbrace{\ket{0}_{\mathrm{C}}\ket{0}_{\mathrm{T}}+\ket{0}_{\mathrm{C}}\ket{1}_{\mathrm{T}}+\ket{1}_{\mathrm{C}}\ket{0}_{\mathrm{T}}}_{\text{block}}-\underbrace{\ket{1}_{\mathrm{C}}\ket{1}_{\mathrm{T}}}_{\text{pass}}\bigr),
\label{eq: after H_1}
\end{align}
where $\ket{0}_{\mathrm{x}}$ shows the presence of the particle(s) and $\ket{1}_{\mathrm{x}}$ shows the absence of the particle(s); and $\mathrm{x}\in\lbrace\mathrm{C},\mathrm{T}\rbrace$. The ancillary photon undergoes $\pi$ rotation only for $\ket{1}_{\mathrm{C}}\ket{1}_{\mathrm{T}}$ state. 

To discriminate between the orthonormal Bell states, the ancillary photon need not to be absorbed by the entangled particle(s). In case the ancillary photon is absorbed by the entangled particle(s), the Bell states transform into a set of four non-orthogonal bipartite states as:
\begin{align}
\ket{\Phi^{\pm}}
&
\rightarrow\dfrac{1}{\sqrt{3}}\left(\ket{0}_{\mathrm{C}}\ket{0}_{\mathrm{T}}+\ket{0}_{\mathrm{C}}\ket{1}_{\mathrm{T}}\pm\ket{1}_{\mathrm{C}}\ket{0}_{\mathrm{T}}\right), \label{eq: non-orthogonal 1}
\\
\ket{\Psi^{\pm}}
&
\rightarrow\dfrac{1}{\sqrt{3}}\left(\ket{0}_{\mathrm{C}}\ket{0}_{\mathrm{T}}-\ket{0}_{\mathrm{C}}\ket{1}_{\mathrm{T}}\pm\ket{1}_{\mathrm{C}}\ket{0}_{\mathrm{T}}\right). \label{eq: non-orthogonal 2}
\end{align}
Using the fact from~\eqref{eq: after H_1} that the probability that the QZ gate is in blocking state is equal to $3/4$ (same for all Bell states), we can obtain the nonabsorbing probability~\eqref{eq:5} and the throughput efficiency~\eqref{eq:6} for the QZ Bell-state analyzer.

\end{document}